\PassOptionsToPackage{numbers,compress,super}{natbib}
\documentclass[pdflatex, sn-nature, iicol, breaklinks=true]{sn-jnl} 

\usepackage{graphicx} 
\usepackage{dcolumn}
\usepackage{bm}
\usepackage{multirow}
\usepackage{booktabs}
\usepackage{amsmath,amssymb,amsfonts}%

\usepackage{times}
\usepackage{titlesec}
\titleformat{\section}
  {\normalfont\normalsize\bfseries}
  {\thesection}{1em}{}
\titleformat{\subsection}
  {\normalfont\normalsize\bfseries}
  {\thesubsection}{1em}{}
\titleformat{\subsubsection}
  {\normalfont\normalsize\bfseries}
  {\thesubsubsection}{1em}{}

\newcommand{\ket}[1]{\lvert#1\rangle} 
\unnumbered 

\begin{document}
\title[Article Title]{Approximate Amplitude Encoding with the Adaptive Interpolating Quantum Transform}

\author*[1,2]{\fnm{Gekko} \sur{Budiutama}}\email{bgekko@quemix.com}
\author*[3]{\fnm{Shunsuke} \sur{Daimon}}\email{daimon.shunsuke@qst.go.jp}
\author[1,2]{\fnm{Xinchi} \sur{Huang}}
\author[1,2]{\fnm{Hirofumi} \sur{Nishi}}
\author[1,2,3]{\fnm{Yu-ichiro} \sur{Matsushita}}

\affil[1]{\orgname{Quemix Inc.}, \orgaddress{\street{Taiyo Life Nihonbashi Building, 2-11-2 Nihonbashi}, \city{Chuo-ku}, \state{Tokyo}, \postcode{103-0027}, \country{Japan}}}
\affil[2]{\orgdiv{Department of Physics}, \orgname{The University of Tokyo}, \orgaddress{\state{Tokyo}, \postcode{113-0033}, \country{Japan}}}
\affil[3]{\orgname{Quantum Materials and Applications Research Center, National Institutes for Quantum Science and Technology (QST)}, \orgaddress{\state{Tokyo}, \postcode{152-8550}, \country{Japan}}}

\abstract{Amplitude encoding of real-world data on quantum computers is often the workflow bottleneck: direct amplitude encoding scales poorly with input size and can offset any speedups in subsequent processing. Fourier-based sparse amplitude encoding lowers cost by retaining only a small subset of dominant coefficients, but its fixed, non-adaptive basis leads to significant information loss. In this work, we replace the Fourier transform with the adaptive interpolating quantum transform (AIQT) in the sparse amplitude encoding workflow. The AIQT learns a data-adapted basis that concentrates information into a small number of coefficients. Consequently, at matched sparsity, the AIQT retains more information and achieves lower reconstruction error compared to the Fourier baseline. On financial time-series data, the AIQT reduces reconstruction error by 40\% relative to the Fourier baseline, and on image datasets the reduction is up to 50\% at the same sparsity level, with nearly identical encoding gate cost. Crucially, the approach preserves the efficiency of Fourier-based methods: the AIQT is built on the structure of the quantum Fourier transform circuit. Its gate count scales quadratically with the number of qubits, while classical evaluation can be carried out in quasilinear time. In addition, the AIQT is trained without labels and does not require sampling from quantum hardware or a simulator, removing a major bottleneck in data-driven amplitude-encoding methods.
}

\maketitle

Quantum computing has made steady progress toward practical applications in areas such as quantum chemistry, optimization, and quantum machine learning \cite{nielsen2010, opening_2, opening_3, opening_4, opening_5, opening_6, opening_7, opening_8, budiutama2024}. In these applications, encoding classical data onto a quantum device is a prerequisite for any end-to-end advantage.

Amplitude encoding is a widely used method that maps a length-$N$ vector to the amplitudes of the computational-basis states of an $n$-qubit system ($n=\log_2 N$). Generic amplitude encoding methods require either an exponential number of two-qubit gates or an exponential number of ancilla qubits, which can cancel out any downstream speedup \cite{ae_1, ae_2, ae_3, ae_4, ae_5, ae_6}. Under certain data assumptions, polynomial-time methods exist, for example for structured probability distributions or localized basis expansions, but their practical effectiveness is often curtailed by device noise and calibration limitations \cite{ae_7, ae_8, ae_9, ae_10, ae_11}. NISQ-friendly amplitude-encoding circuits based on data-driven quantum circuit models have also been proposed \cite{ae_13,ae_14,ae_12,ae_15,ae_16}, but they typically rely on explicit quantum circuit evaluation on hardware or high-fidelity simulators, either during variational training or for large-scale circuit synthesis.

One practical way to lower the cost of amplitude encoding is to prepare an approximate state by extracting dominant coefficients \cite{topkFourier_theory, topkFourier_app, hwsl}. In this workflow, a classical transform (e.g., Fourier or Walsh-Haar) is applied to the input to obtain its coefficients in the transform basis. The dominant coefficients are then selected and the rest discarded. Amplitude encoding is applied to the retained coefficients, and the original vector is approximately reconstructed via the inverse transform. However, this sparse encoding workflow is inherently lossy: the use of a fixed transform across datasets can drop non-dominant yet task-relevant details. For example, edges, textures, and transients are not compactly represented in a Fourier basis, which can raise reconstruction error for images or non-smooth time series. While methods to improve the expressiveness of fixed transforms exist \cite{gt}, it is not clear how to translate these learned transforms into compact, hardware-efficient quantum circuits.

To address this limitation, we apply the adaptive interpolating quantum transform (AIQT) to sparse amplitude encoding. Introduced in our prior work \cite{aiqt}, the AIQT is a quantum-native, unitary, compactly parameterized transform that adapts its basis to the data. The AIQT is trained to maximize the information captured by the retained coefficients, reducing information loss in the sparse selection step. Empirically, at matched sparsity, the AIQT outperforms Fourier baselines, reducing reconstruction error by 40\% on 1D signals and up to 50\% on 2D images. Unlike prior data-driven amplitude-encoding methods, our training pipeline operates entirely on classical data and does not require sampling from a quantum device or simulator at any stage. Moreover, since the AIQT is built on the efficient circuit structure of the quantum Fourier transform, its gate count scales polynomially (quadratically) with the number of qubits.

\begin{figure*}[t]
    \centering
    \includegraphics[width=0.9\textwidth]{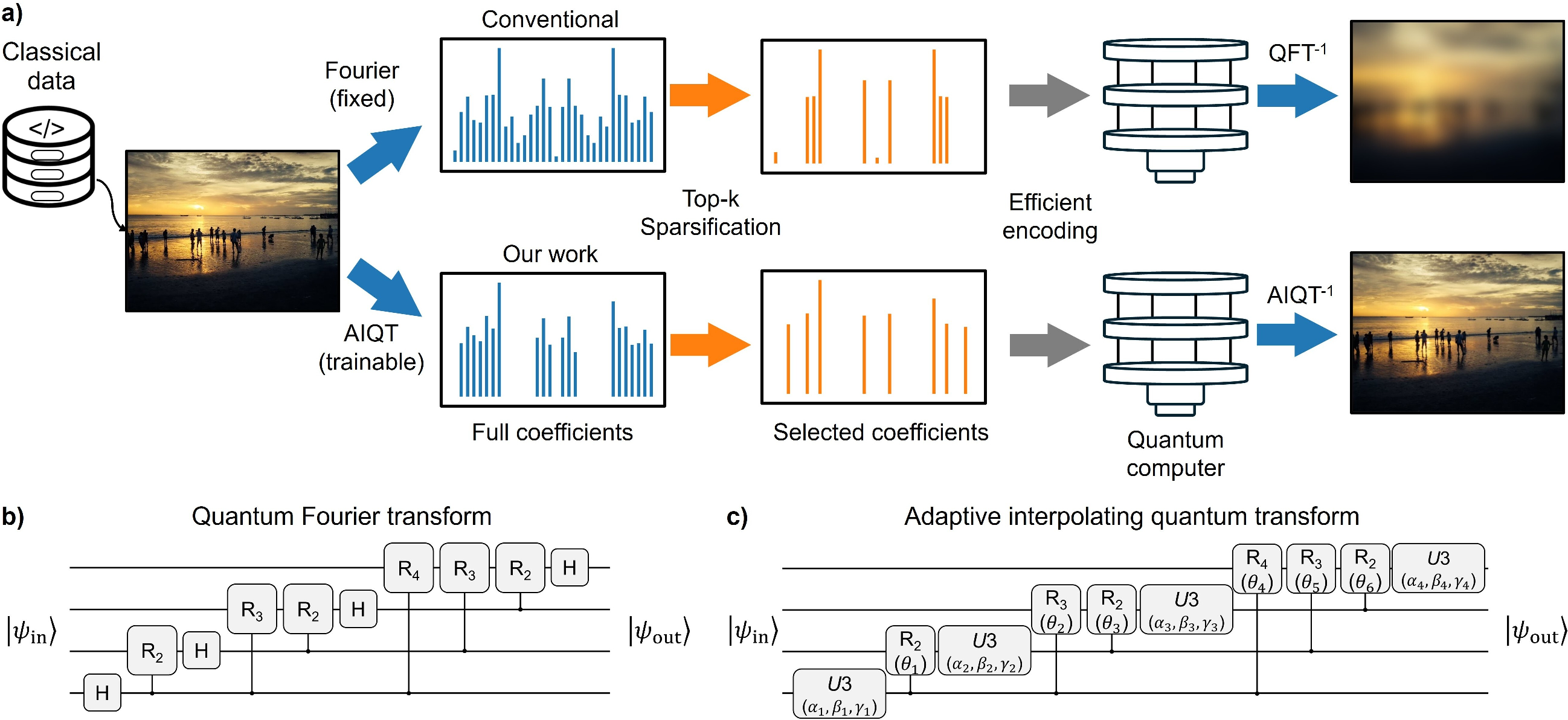}
    \caption{
    \textbf{Sparse amplitude encoding workflows with Fourier and AIQT transforms and their quantum circuits.} a) In sparse amplitude encoding, a transform (Fourier transform in the upper panel or AIQT in the lower panel) is first applied to the classical data $\mathbf{x}$, producing the full set of coefficients $\mathbf{y}$. We then keep the $k$ largest-magnitude coefficients and set the rest to zero, yielding a $k$-sparse vector $\tilde{\mathbf{y}}$. This vector is amplitude encoded on the quantum computer to obtain the state $\ket{\tilde{\phi}}$. Finally, an approximate reconstruction of the data is obtained by applying the corresponding inverse transform, producing the state $\ket{\tilde{\psi}}$. The AIQT is trained to maximize the amount of information preserved during coefficient truncation, leading to lower reconstruction error. b) The quantum Fourier transform (QFT) circuit used in FSL.
    c) The AIQT circuit used in the proposed method. The circuit is constructed by parameterizing the QFT structure with single-qubit gates $\mathrm{U3}(\alpha,\beta,\gamma)$ and controlled-phase gates $\mathrm{CR}(\theta)$, yielding $U_{\mathrm{AIQT}}(\boldsymbol{\theta},\boldsymbol{\alpha},\boldsymbol{\beta},\boldsymbol{\gamma})$. Here, $\mathrm{U3}(\alpha,\beta,\gamma)
    =
    \bigl[\begin{smallmatrix}
    \cos(\alpha/2) & -e^{i\gamma}\sin(\alpha/2)\\
    e^{i\beta}\sin(\alpha/2) & e^{i(\beta+\gamma)}\cos(\alpha/2)
    \end{smallmatrix}\bigr]$
    and $\mathrm{CR}(\theta) = \mathrm{diag}(1, 1, 1, e^{i\theta}).$ 
}
    \label{fig1}
\end{figure*}

\section{Transform-based Approximate Amplitude Encoding}
\label{sec:workflow}
The general workflow for transform-based approximate amplitude encoding proceeds as follows. For an input $\mathbf{x} \in \mathbb{R}^N$, a unitary transform $U$ is first applied on a classical computer to obtain complex coefficients $\mathbf{y}(\mathbf{x}) = U \mathbf{x} \in \mathbb{C}^N.$ Let $\mathcal{K}(\mathbf{x})$ index the $k$ coefficients of $\mathbf{y}(\mathbf{x})$ with largest squared magnitude. The corresponding sparse coefficient vector $\tilde{\mathbf{y}}(\mathbf{x}) \in \mathbb{C}^N$ is defined by
\begin{equation}
    \tilde{y}_j(\mathbf{x})
    =
    \begin{cases}
        y_j(\mathbf{x}), & j \in \mathcal{K}(\mathbf{x}),\\[3pt]
        0,               & j \notin \mathcal{K}(\mathbf{x}).
    \end{cases}
\end{equation}
The normalized truncated coefficient is then defined as
\begin{equation}
    \phi_j(\mathbf{x})
    =
    \frac{\tilde{y}_j(\mathbf{x})}{
        \left(
            \sum_{\ell=1}^N \bigl|\tilde{y}_\ell(\mathbf{x})\bigr|^2
        \right)^{1/2}
    }.
\end{equation}
On the quantum device, a sparse state-preparation circuit encodes these
coefficients into an $n = \log_2 N$ qubit state,
\begin{equation}
    \ket{\phi(\mathbf{x})}
    =
    \sum_{j=1}^N \phi_j(\mathbf{x})\,\ket{j}.
\end{equation}
Finally, an inverse transform $U^{-1}$ is implemented as a quantum circuit,
yielding an approximate reconstruction of the original vector,
\begin{equation}
    \ket{\tilde{\psi}(\mathbf{x})} = U^{-1}\ket{\phi(\mathbf{x})}.
\end{equation}
Here, $\ket{\tilde{\psi}(\mathbf{x})}$ denotes the reconstructed state that approximates the exact amplitude-encoded state $\ket{\psi(\mathbf{x})}$, where $\ket{\psi(\mathbf{x})}=\frac{1}{\|\mathbf{x}\|_2}\sum_{j=1}^{N} x_j \ket{j}$.

Several approximate state-preparation methods fit into this workflow. The first proposal uses the Fourier transform and refers to this scheme as the Fourier series loader (FSL)~\cite{topkFourier_theory}. More recent work instantiates the same hybrid template with the Walsh-Haar transform in place of the Fourier transform, improving the scaling of the inverse transform while otherwise following the same overall workflow~\cite{hwsl}. In this work we consider the Fourier-based instance, the FSL, to be our baseline transform-based approximate amplitude-encoding scheme.

From a resource point of view, the FSL workflow consists of a classical
preprocessing step to compute $\mathbf{y}(\mathbf{x})$ and select the top-$k$ coefficients, followed by sparse amplitude encoding and an inverse transform on the quantum device. Among sparse state-preparation methods, the worst-case circuit size scales as $O\!\big(\tfrac{n k}{\log n} + n\big)$~\cite{sparse1, sparse5, sparse2, sparse3, sparse4, mao_sparse_encode, li_sparse_encode}. Thus, the end-to-end cost for the FSL is $O(N\log N)$ on the classical side for preprocessing (dominated by the FFT) and $O\!\big(\tfrac{n k}{\log n}+n+n^{2}\big)$ on the quantum side for sparse amplitude encoding followed by the inverse transform. In regimes where the target states admit good $k$-sparse approximations with $k \ll N$, this method uses fewer quantum gates than generic amplitude-encoding schemes, whose worst-case circuit size scales linearly in $N$ for arbitrary states. 

\begin{figure}[!t]
    \centering
    \includegraphics[width=0.47\textwidth]{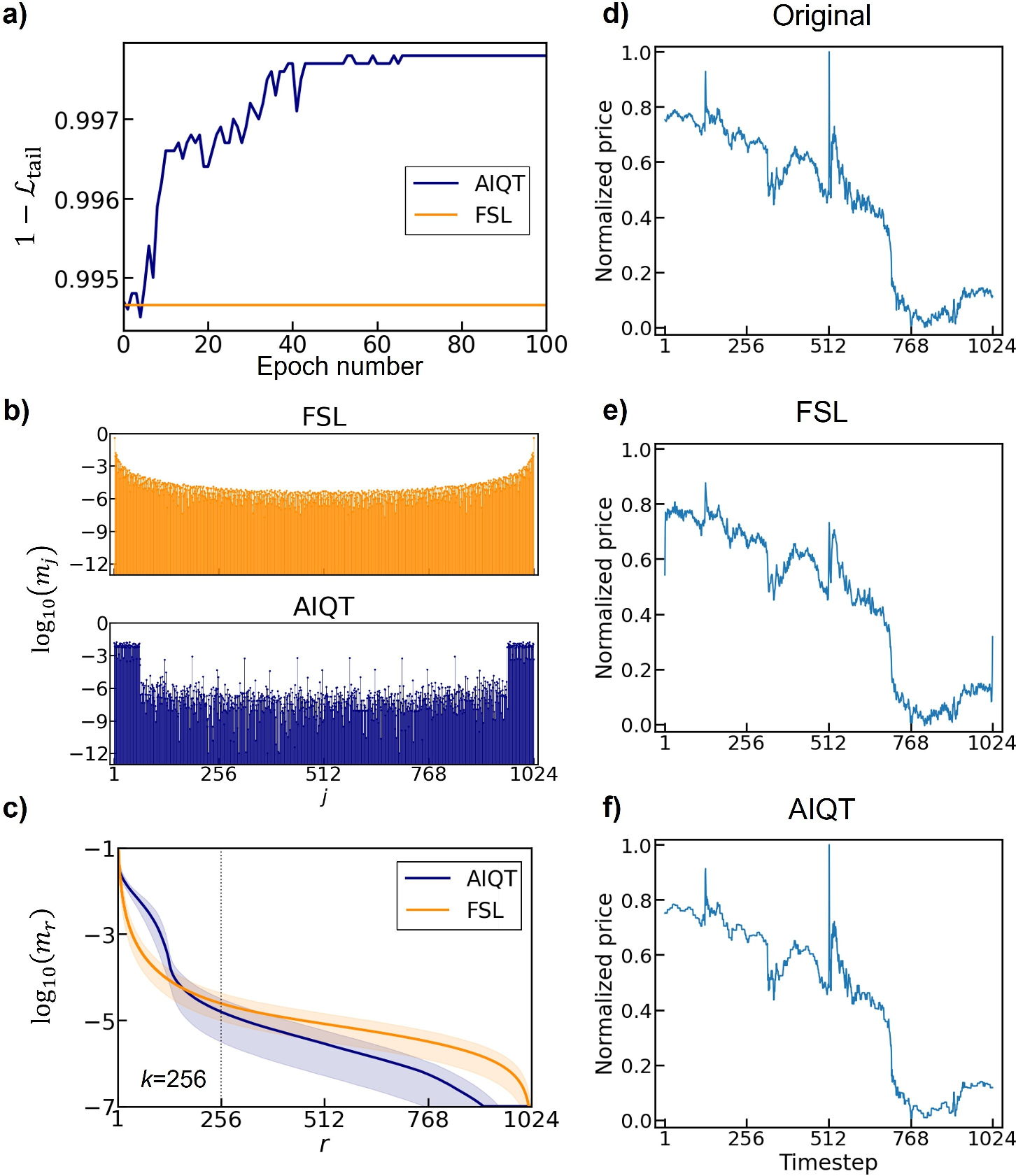}
    \caption{\textbf{Coefficient statistics and single-sample reconstructions for the AIQT and the FSL.} a) Dataset-averaged squared magnitude of the top-256 coefficients as a function of training epoch for the AIQT (blue), compared with the fixed FSL baseline (orange). b) Distribution of squared magnitudes ($m_j$) across coefficient indices $j \in \{1,\dots,1024\}$ for a representative sample, shown for the FSL (top, orange) and the AIQT (bottom, blue). c) Average squared magnitude ($m_r$) as a function of rank $r$ for the AIQT (blue) and the FSL (orange). For each input, coefficients are sorted in descending order, from largest ($m_1$) to smallest ($m_{1024}$), and $m_r$ is then averaged over the dataset. Solid lines show the dataset mean, and shaded regions indicate the standard deviation. d-f) Single-sample visualization showing the original signal (d), the FSL reconstruction (e), and the AIQT reconstruction (f), all at matched sparsity ($k=256$).}
    \label{fig2}
\end{figure}

\section{The Adaptive Interpolating Quantum Transform for Amplitude Encoding}
The adaptive interpolating quantum transform (AIQT) is a framework for constructing trainable unitary transforms that interpolate among a set of predefined quantum transforms. In general, the AIQT framework is circuit agnostic: it can be instantiated using different efficient quantum circuits, with interpolation achieved by embedding trainable parameters into them \cite{aiqt}.

For this work, we utilized a parametrized variant of the quantum Fourier transform (Figs.~\ref{fig1}b and c). Specifically, we replace each Hadamard and single-qubit phase gate by a generic single-qubit gate $\mathrm{U3}(\alpha,\beta,\gamma) = \bigl[\begin{smallmatrix} \cos(\alpha/2) & -e^{i\gamma}\sin(\alpha/2)\\ e^{i\beta}\sin(\alpha/2) & e^{i(\beta+\gamma)}\cos(\alpha/2) \end{smallmatrix}\bigr]$, and parametrize the controlled phase gates as $\mathrm{CR}(\theta) = \mathrm{diag}(1, 1, 1, e^{i\theta})$. With this parameterization, the AIQT smoothly interpolates between the quantum Fourier transform, the Hadamard transform, and the identity as its parameters vary. We choose this AIQT for two reasons: (i) like the FFT, this AIQT admits a butterfly-style factorization, yielding an $O(N\log N)$ classical algorithm; and (ii) the corresponding quantum implementation on $n=\log_2 N$ qubits has a gate count of $O(n^{2})$.

Figure~\ref{fig1}a shows that the AIQT-based approximate amplitude-encoding workflow follows the same transform-based pipeline as the FSL baseline. For a real input $\mathbf{x}\in\mathbb{R}^N$, applying the AIQT unitary $U_{\mathrm{AIQT}}(\boldsymbol{\theta},\boldsymbol{\alpha},\boldsymbol{\beta},\boldsymbol{\gamma})$ yields complex transform coefficients $\mathbf{y}(\mathbf{x}) = U_{\mathrm{AIQT}}\,\mathbf{x}$. The sparse coefficient vector $\tilde{\mathbf{y}}(\mathbf{x})$ is obtained by retaining only those entries whose indices are in $\mathcal{K}(\mathbf{x})$, where $\mathcal{K}(\mathbf{x})$ denotes the index set of the $k$ largest squared-magnitude coefficients of $\mathbf{y}(\mathbf{x})$, and setting all remaining entries to zero. This sparse vector is amplitude encoded on the quantum computer to obtain $\ket{\phi(\mathbf{x})}$, and an approximate reconstruction of $\mathbf{x}$ is recovered by applying the inverse transform
$U_{\mathrm{AIQT}}^{-1}$.

To train the AIQT, we first define the normalized squared magnitudes
\begin{equation}
    m_j(\mathbf{x})
    =
    \frac{\lvert y_j(\mathbf{x}) \rvert^2}{
        \sum_{\ell=1}^N \lvert y_\ell(\mathbf{x}) \rvert^2
    },
    \label{eq:mj}
\end{equation}
so that $\sum_j m_j(\mathbf{x})=1$ for each input $\mathbf{x}$. Here, a top-$k$ tail loss function is used to train the AIQT to encourage the transform to concentrate information in the top-$k$ largest coefficients:
\begin{equation}
\label{eq:tail}
\mathcal{L}_{\text{tail}}
=
\mathbb{E}_{\mathbf{x}} \left[
    \sum_{j \notin \mathcal{K}(\mathbf{x})} m_j(\mathbf{x})
\right],
\end{equation}
where $\mathbb{E}_{\mathbf{x}}[\cdot]$ denotes the average over the dataset.

We evaluate both AIQT and the FSL baseline by comparing their reconstructed states $\ket{\tilde{\psi}_{\mathrm{AIQT}}(\mathbf{x})}$ and $\ket{\tilde{\psi}_{\mathrm{FSL}}(\mathbf{x})}$ to the exact amplitude-encoded state $\ket{\psi(\mathbf{x})}$, using the complex root-mean-squared error (cRMSE) and the fidelity $F$ (see Methods).

\section{Results}
\subsection{The AIQT for efficient amplitude encoding}
\begin{table}[t]
\centering
\caption{\textbf{Performance of the FSL vs the AIQT across sparsity $k$ on the one-dimensional finance dataset.}}
\label{table1}
\setlength{\tabcolsep}{3pt}
\renewcommand{\arraystretch}{1.15}
\begin{tabular}{r l cc cc}
\toprule
\multirow{2}{*}{$k$} & \multirow{2}{*}{Method}
  & \multicolumn{2}{c}{Train}
  & \multicolumn{2}{c}{Validation} \\
\cmidrule(lr){3-4} \cmidrule(lr){5-6}
 & & cRMSE $\times 10^{-3}$ & $F$
   & cRMSE $\times 10^{-3}$ & $F$ \\
\midrule
\multirow{2}{*}{64}  & AIQT & 4.144 & 0.9826 & 4.197 & 0.9821 \\                     & FSL  &  ---  &  ---   & 4.659 & 0.9780 \\
\midrule
\multirow{2}{*}{128} & AIQT &  2.462 & 0.9938 &  2.468 & 0.9938 \\
                     & FSL  &   ---  &  ---   & 3.308 & 0.9888 \\
\midrule
\multirow{2}{*}{256} & AIQT &  1.471 & 0.9978 &  1.455 & 0.9978 \\
                     & FSL  &   ---  &  ---   &  2.283 & 0.9947 \\
\midrule
\multirow{2}{*}{384} & AIQT & 1.055 & 0.9989 & 1.036 & 0.9989 \\
                     & FSL  &   ---  &  ---   & 1.731 & 0.9969 \\
\bottomrule
\end{tabular}
\end{table}

Figure~\ref{fig2}a shows the total squared magnitude captured by the top-$k$ coefficients ($1-\mathcal{L}_{\text{tail}}$) with $k=256$ (25\% of $N=1024$). The AIQT (blue) quickly surpasses the FSL (orange, $1-\mathcal{L}_{\text{tail}}\approx 0.995$) and converges to a higher plateau ($1-\mathcal{L}_{\text{tail}}\approx 0.998$), indicating that the AIQT basis consistently concentrates more information into the top-256 set than Fourier.

Figure~\ref{fig2}b shows a representative example of how the squared-magnitude coefficients $m_j$ are distributed across index $j$ for the FSL (top, orange) and the AIQT (bottom, blue). In this example, the FSL spreads the squared magnitudes $m_j$ over many Fourier modes, with an almost uniform background level except for the two dominant coefficients at $j=2$ and its conjugate-symmetric pair at $j=1024$. In contrast, the AIQT (bottom, blue) concentrates more information into the first and last $\sim 50$ bins and into a small number of pronounced coefficients in the middle of the spectrum, while pushing the remaining coefficients down to a much lower background level. This visualization shows how the AIQT reshapes the input so that as much information as possible is concentrated into the top-$k$ coefficients, reducing the loss incurred during the truncation step.

Figure~\ref{fig2}c shows the dataset-averaged squared magnitude $m_r$ as a function of rank $r$, where coefficients are ranked in descending order of squared magnitude for each input, with $r=1$ corresponding to the largest coefficient and $r=1024$ to the smallest. The vertical black dotted line marks $r=k$. Compared to the FSL (orange), the AIQT (blue) concentrates more information to the leading ranks and less to the tail: the blue curve lies above the orange curve for $r \le k$ and below for $r>k$, indicating a steeper decay beyond the kept coefficients. Shaded bands show across-sample variability and confirm that this shift of coefficient magnitude toward the top-$k$ coefficients is systematic rather than due to a few outliers.

Figures~\ref{fig2}d--f show a single-sample comparison: (d) the original time series, (e) the FSL reconstruction using the top-256 Fourier coefficients, and (f) the AIQT reconstruction using the top-256 AIQT coefficients. Under the same 256-coefficient budget, both reconstructions follow the overall trend, but the AIQT result better preserves turning points and local variations. For instance, the AIQT recovers a prominent peak that is noticeably smoothed out in the FSL reconstruction. This qualitative behavior is consistent with the higher information capture in Fig.~\ref{fig2}a and the steeper rank profile in Fig.~\ref{fig2}c.

Table~\ref{table1} compares the FSL and the AIQT across sparsity levels $k$ on the one-dimensional finance dataset. The FSL is a fixed (non-trainable) Fourier transform, so its entries are reported only on the validation split, whereas the AIQT is trained on the training split and evaluated on both splits. Across all $k$, the AIQT achieves lower cRMSE and higher fidelity than that of the FSL. The improvement grows with sparsity: at $k=64$, the AIQT reduces validation cRMSE from $4.659$$\times10^{-3}$ to $4.197$$\times10^{-3}$, a reduction of roughly 10\%, while at $k=384$ the validation cRMSE drops from $1.731$$\times10^{-3}$ to $1.036$$\times10^{-3}$, about a 40\% reduction, with fidelity increasing from $0.9969$ to $0.9989$. 

\begin{figure}[!t]
    \centering
    \includegraphics[width=0.40\textwidth]{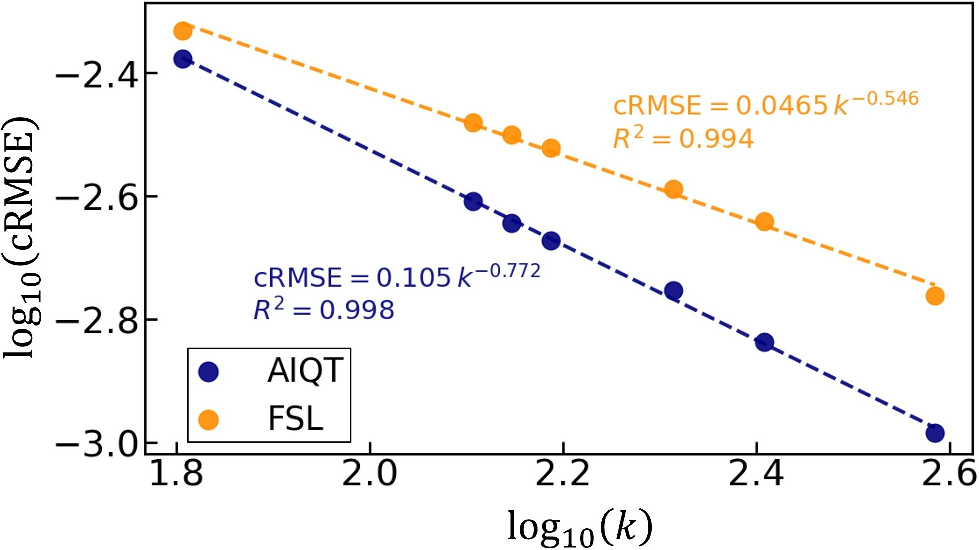}
    \caption{\textbf{Reconstruction error as a function of the number of retained coefficients.} The complex root-mean-squared error (cRMSE) in relation to the number of retained coefficients ($k$) for the FSL and the AIQT on the finance dataset. Dashed lines denote power-law fits ($\mathrm{cRMSE}=Ak^B$) obtained by ordinary least squares on log-transformed data. The coefficient of determination ($R^2$) denotes the goodness of fit for the linearized model in log space.}
    \label{fig3}
\end{figure}

Figure~\ref{fig3} shows the relationship between reconstruction error (cRMSE) and the number of retained coefficients ($k$) for the FSL (orange) and the AIQT (blue). As discussed, $k$ directly determines the cost of sparse amplitude encoding on a quantum computer: preparing a $k$-sparse state has worst-case circuit size scaling as $O\!\big(\tfrac{n k}{\log n}+n\big)$~\cite{sparse1, sparse5, sparse2, sparse3, sparse4, mao_sparse_encode, li_sparse_encode}. Compared to the FSL, the error decays more rapidly with $k$ for the AIQT, with $\mathrm{cRMSE}\propto k^{-0.772}$ rather than that of the FSL ($\mathrm{cRMSE}\propto k^{-0.546}$). Thus, on the finance dataset, the AIQT does not merely improve upon the FSL by a constant factor, but instead improves the scaling of the workflow, becoming increasingly advantageous as higher reconstruction accuracy (higher $k$) is required.

\subsection{Imaginary residuals in the AIQT reconstructions}
As our goal is to encode real datasets as amplitudes on a quantum computer, we must be careful about any imaginary components that might arise in either the FSL or AIQT pipeline.

For the FSL, the Fourier spectrum of a real-valued input is conjugate-symmetric; therefore, as long as the truncation preserves this symmetry, the reconstructed state is purely real. In contrast, the AIQT operates in the complex domain and does not enforce this conjugate-symmetry structure, so reconstructed states can, in principle, have non-zero imaginary components even when the input is real. To quantify this imaginary leakage in the reconstruction, we define the per-sample imaginary-part norm
\begin{equation}
    I(\mathbf{x})
    =
    \big\|\operatorname{Im}\big(\tilde{\boldsymbol{\psi}}(\mathbf{x})\big)\big\|_2
    =
    \left(
        \sum_{j=1}^N
        \big(\operatorname{Im}\big(\tilde{\psi}_j(\mathbf{x})\big)\big)^2
    \right)^{1/2}.
\end{equation} We then consider its dataset average
\begin{equation}
    \overline{I}
    =
    \mathbb{E}_{\mathbf{x}}\big[ I(\mathbf{x}) \big].
\end{equation}

For comparison, we also track the real part of the reconstruction. We define the per-sample real-part norm
\begin{equation}
    R(\mathbf{x})
    =
    \big\|\operatorname{Re}\big(\tilde{\boldsymbol{\psi}}(\mathbf{x})\big)\big\|_2
    =
    \left(
        \sum_{j=1}^N
        \big(\operatorname{Re}\big(\tilde{\psi}_j(\mathbf{x})\big)\big)^2
    \right)^{1/2},
\end{equation} and its dataset average
\begin{equation}
    \overline{R}
    =
    \mathbb{E}_{\mathbf{x}}\big[ R(\mathbf{x}) \big].
\end{equation}

\begin{figure}[!t]
    \centering
    \includegraphics[width=0.40\textwidth]{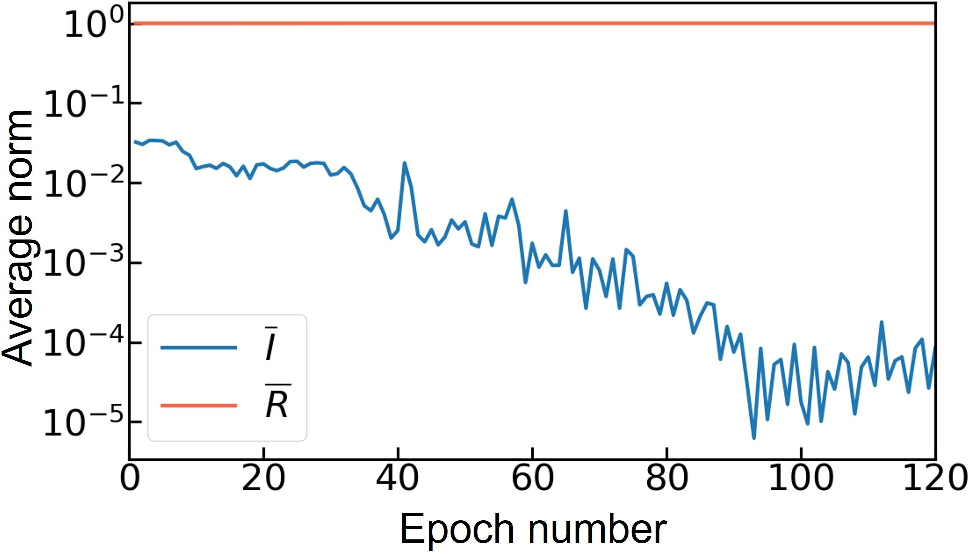}
    \caption{\textbf{Real and imaginary norms of the AIQT reconstruction during training.} Dataset-averaged imaginary-part norm ($\overline{I}$, blue) and dataset-averaged real-part norm ($\overline{R}$, red) of the AIQT reconstructed signal as a function of training epoch for the AIQT model with $k=256$. The model automatically reduces leakage into the imaginary part over training.}
    \label{fig4}
\end{figure}

Figure~\ref{fig4} shows the evolution of the dataset-averaged norms $\overline{I}$ (blue) and $\overline{R}$ (red) of the reconstructed state for $k=256$ on a logarithmic scale. Over training, $\overline{I}$ drops by roughly three orders of magnitude, from about $10^{-2}$ to $10^{-5}$, while $\overline{R}$ remains essentially pinned at $1$. This indicates that the AIQT learns parameters that keep almost all of the amplitude norm in the real part and drive the imaginary leakage to a negligible level.

Table~\ref{table2} complements this plot by reporting the average real- and imaginary-part norms across sparsity levels $k$ on the finance dataset. For each $k$ we list $1 - \overline{R}$, which measures how far the average real-part norm $\overline{R}$ deviates from $1$, together with the imaginary-part norm $\overline{I}$, for both train and validation splits. Except at the lowest sparsity ($k=64$), the deviation $1 - \overline{R}$ is on the order of $10^{-9}$ and $\overline{I}$ is on the order of $10^{-4}$ to $10^{-5}$, indicating that the reconstructed states are almost perfectly real-valued and that imaginary leakage remains negligible in practice. The close agreement between train and validation values further suggests that this behavior is stable and not an artifact of overfitting.

\begin{table}[t]
\centering
\caption{\textbf{Real- and imaginary-part residual norms for the AIQT across sparsity $k$ on the one-dimensional finance dataset.}}
\label{table2}
\setlength{\tabcolsep}{3.5pt}
\renewcommand{\arraystretch}{1.15}
\begin{tabular}{r cccc}
\toprule
\multirow{2}{*}{$k$}
  & \multicolumn{2}{c}{Train}
  & \multicolumn{2}{c}{Validation} \\
\cmidrule(lr){2-3} \cmidrule(lr){4-5}
 & $1-\overline{R}$ & $\overline{I}$ & $1-\overline{R}$ & $\overline{I}$ \\
\midrule
 64  & 1.387 $\times 10^{-3}$ & 4.896 $\times 10^{-2}$ & 1.439 $\times 10^{-3}$ & 4.975 $\times 10^{-2}$ \\
128  & 1.367 $\times 10^{-9}$ & 5.204 $\times 10^{-5}$ & 1.378 $\times 10^{-9}$ & 5.224 $\times 10^{-5}$ \\
256  & 2.068 $\times 10^{-9}$ & 6.410 $\times 10^{-5}$ & 2.052 $\times 10^{-9}$ & 6.385 $\times 10^{-5}$ \\
384  & 6.148 $\times 10^{-9}$ & 1.106 $\times 10^{-4}$ & 6.156 $\times 10^{-9}$ & 1.107 $\times 10^{-4}$ \\
\bottomrule
\end{tabular}
\end{table}

Thus, while the AIQT lacks the theoretical real-valued guarantee of the FSL, the trained models effectively recover real outputs in practice, so imaginary residuals do not pose a practical limitation for the amplitude-encoding regime considered here.

\subsection{The deep AIQT architecture}

So far we have focused on a single AIQT layer acting as a learned replacement for the Fourier transform. In practice, it can be beneficial to increase the expressive power of the transform by composing several such layers, in direct analogy with adding depth in a neural network.

Here we propose the deep AIQT architecture, which increases the expressive power of the transform by stacking multiple AIQT blocks.

\begin{figure}[!t]
    \centering
    \includegraphics[width=0.45\textwidth]{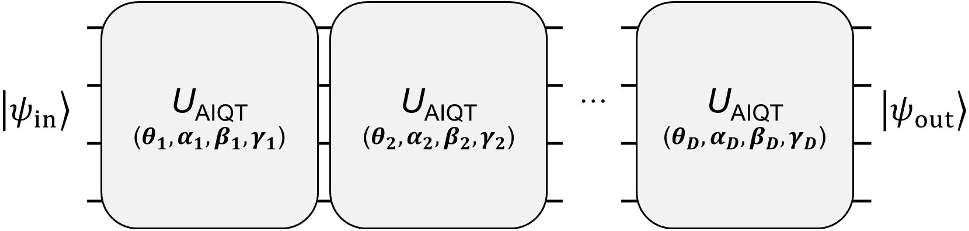}
    \caption{\textbf{The deep AIQT architecture.} A depth-$D$ stack of AIQT blocks with independent parameters, applied sequentially to the input state.
}
    \label{fig5}
\end{figure}

Figure~\ref{fig5} depicts the deep AIQT architecture: a stack of $D$ learnable AIQT blocks applied sequentially to the input state. Let $\boldsymbol{\eta}_d =(\boldsymbol{\theta}_d,\boldsymbol{\alpha}_d,\boldsymbol{\beta}_d,\boldsymbol{\gamma}_d)$ denote the parameters of block $d$. The overall unitary is
\begin{equation}
    U_{\text{DeepAIQT}}
    =
    \prod_{d=1}^{D}
    U_{\text{AIQT}}(\boldsymbol{\eta}_d),
\end{equation}
where the factors are ordered so that $d=1$ acts first on the input state. This construction retains the unitarity and FFT-like structure of a single AIQT layer, but increases expressivity by composing multiple trainable transforms. Because each block has $O(N \log N)$ classical cost and $O(n^2)$ quantum gate count (with $N = 2^n$), the overall deep AIQT scales linearly in the number of layers $D$, providing a principled way to trade computation for representational power.

\begin{figure}[!t]
    \centering
    \includegraphics[width=0.48\textwidth]{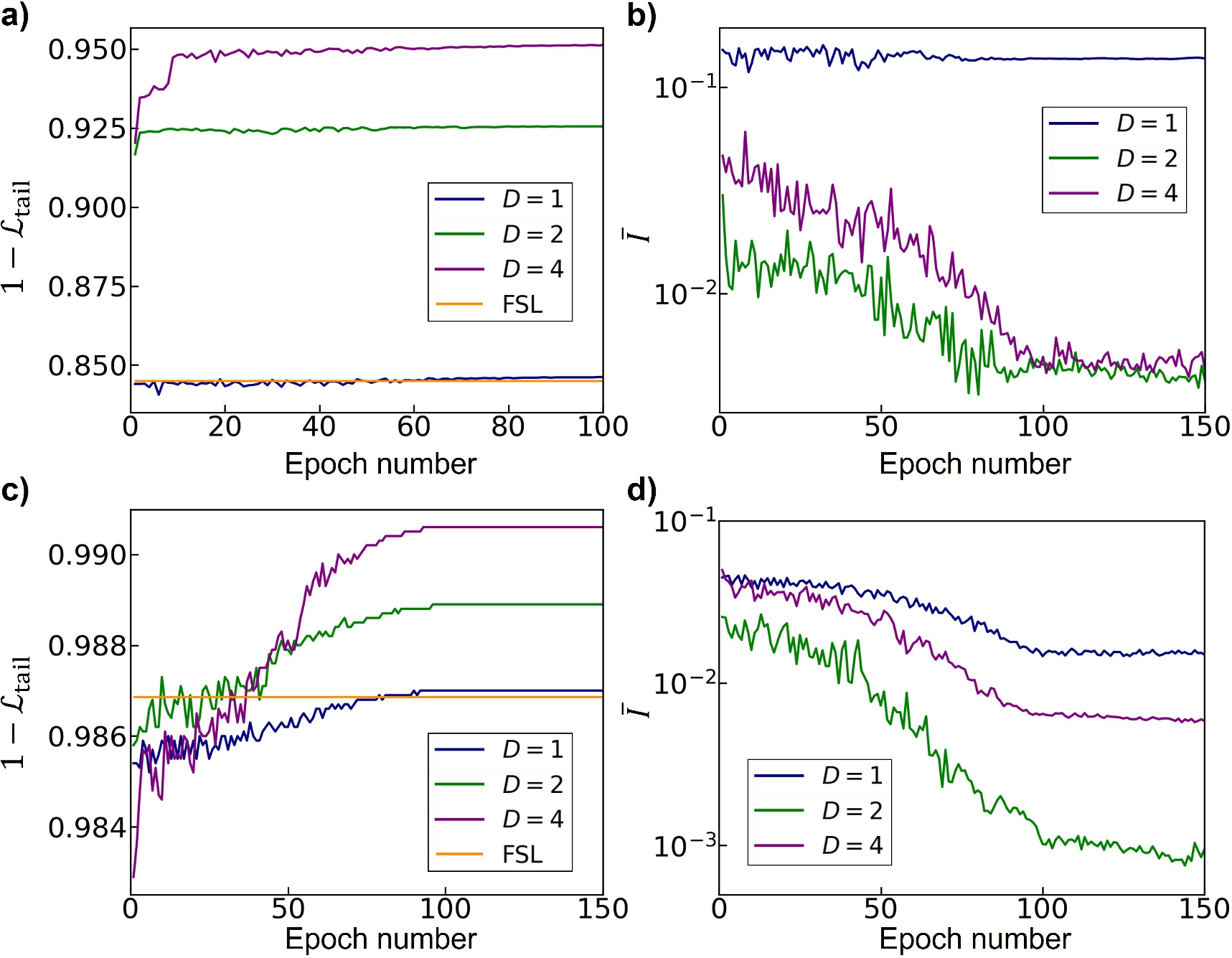}
    \caption{\textbf{Training dynamics of the deep AIQT on MNIST and CIFAR-RGB.} 
    a,b) MNIST results for depths $D\in\{1,2,4\}$, showing the squared magnitude of the dominant coefficients and the average imaginary-part norm. c,d) CIFAR-RGB results in the same format. Sparsity is fixed at $k=52$ for MNIST and $k=154$ for CIFAR-RGB.
}
    \label{fig6}
\end{figure}

Figure~\ref{fig6} summarizes the training dynamics of the deep AIQT for MNIST (top row) and CIFAR-RGB (bottom row). Figures~\ref{fig6}a and c show the total squared magnitude captured by the top-$k$ coefficients, while Fig. 6b and d show the dataset-averaged imaginary-part norm $\overline{I}$ on a logarithmic scale. In both datasets, all AIQT configurations outperform the FSL (orange) in terms of the square magnitude concentration, and increasing the depth from $D=1$ (blue) to $D=4$ (purple) yields a significant improvement (Figs.~\ref{fig6}a and c). For MNIST, deeper models also significantly suppress imaginary leakage: $\overline{I}$ decreases by more than an order of magnitude during training, with $D=2$ (green) and $D=4$ (purple) converging to much smaller values than $D=1$ (Fig.~\ref{fig6}b). On CIFAR-RGB, all depths reduce $\overline{I}$ relative to initialization, and the intermediate depth $D=2$ (green) achieves the smallest imaginary residual (Fig.~\ref{fig6}d). In all configurations, the final values of $\overline{I}$ are small enough that imaginary leakage is negligible for the approximate amplitude-encoding regime considered here.

\begin{figure}[t]
    \centering
    \includegraphics[width=0.45\textwidth]{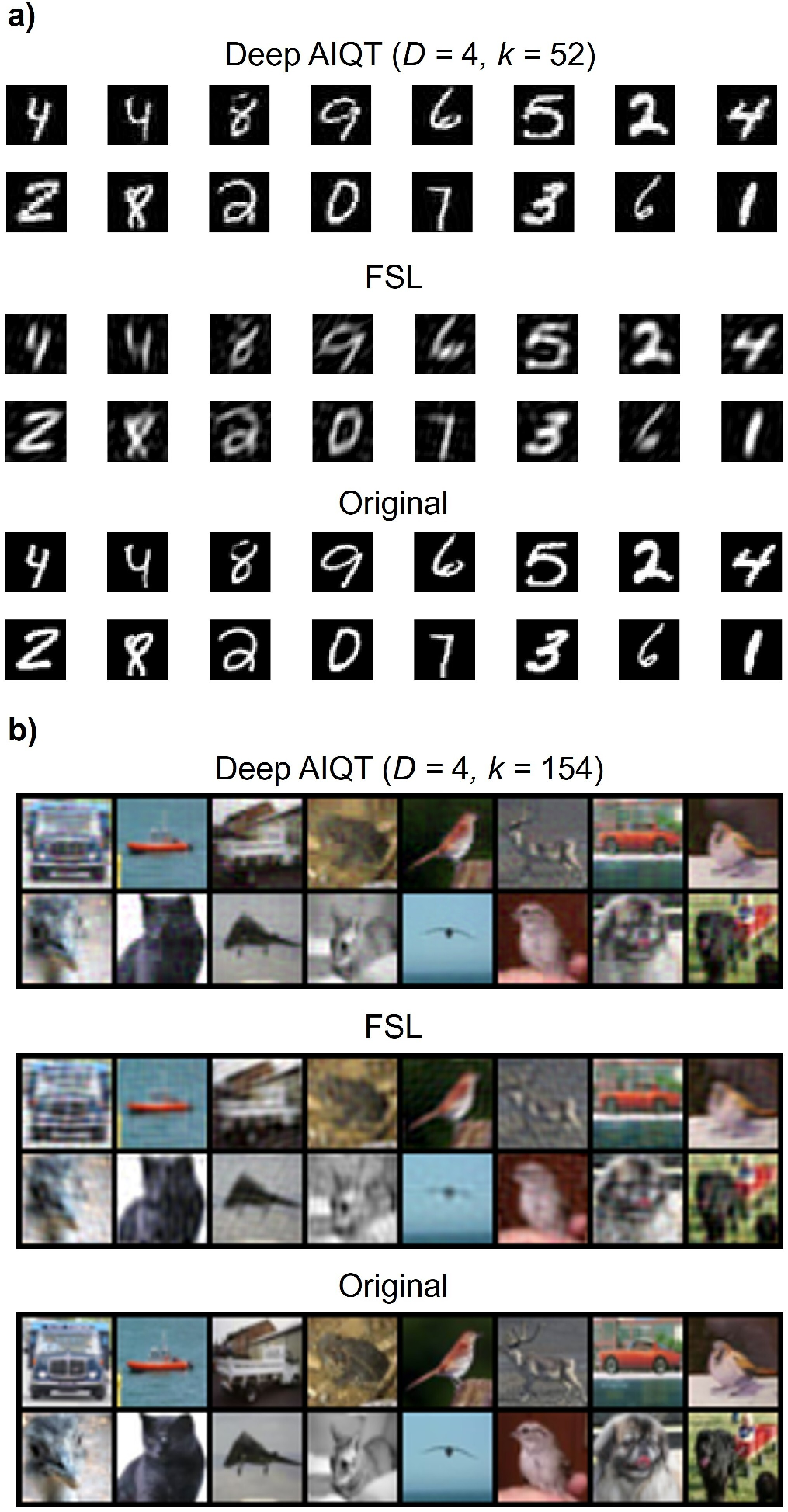}
    \caption{\textbf{Image reconstructions at matched sparsity using the deep AIQT and the FSL.} a) MNIST reconstructions at $k=52$, comparing the deep AIQT ($D=4$) with the FSL. b) CIFAR-RGB reconstructions at $k=154$, comparing the deep AIQT ($D=4$) with the FSL. In each panel, the bottom row shows the original images.
}
    \label{fig7}
\end{figure}

Figure~\ref{fig7}a compares MNIST reconstructions at matched sparsity $k=52$ using the deep AIQT with depth $D=4$. Relative to the FSL, the deep AIQT produces sharper digit strokes, cleaner junctions, and fewer directional artifacts, yielding images that are visually closer to the reference rows. The FSL reconstructions exhibit noticeable blurring around edges.


Figure~\ref{fig7}b shows the same comparison on CIFAR at $k=154$ using the deep AIQT with depth $D=4$. With the same coefficient budget, the deep AIQT better preserves object contours (e.g., vehicle edges and bird outlines), local textures (such as fur and feathers), and smooth color transitions, whereas the Fourier reconstructions appear washed out and lose fine detail.

Table~\ref{table3} reports the performance of the deep AIQT on MNIST and CIFAR at fixed sparsity ($k=52$ for MNIST, $k=154$ for CIFAR), comparing different depths $D$ to the FSL baseline. On all three datasets, increasing depth from $D=1$ to $D=4$ progressively reduces cRMSE and increases fidelity on both training and validation splits. For MNIST, the deepest model ($D=4$) nearly halves the validation cRMSE relative to the FSL (6.934$\times 10^{-3}$ vs.\ 12.46$\times 10^{-3}$) and raises the fidelity from 0.8486 to 0.9515. On CIFAR-B\&W and CIFAR-RGB, $D=4$ likewise improves over the FSL, reducing validation cRMSE by about 17\% and 16\%, respectively, while increasing fidelity.

\begin{table*}[t]
\centering
\caption{\textbf{Performance of the deep AIQT ($D\in\{1,2,4\}$) on MNIST and CIFAR.}}
\label{table3}
\setlength{\tabcolsep}{6pt}
\renewcommand{\arraystretch}{1.15}
\begin{tabular}{l r c c c c c}
\toprule
\multirow{2}{*}{Dataset}
& \multirow{2}{*}{$k$}
& \multirow{2}{*}{$D$}
& \multicolumn{2}{c}{Training}
& \multicolumn{2}{c}{Validation} \\
\cmidrule(lr){4-5}\cmidrule(lr){6-7}
& & 
& cRMSE $\times 10^{-3}$ & $F$
& cRMSE $\times 10^{-3}$ & $F$ \\
\midrule
\multirow{4}{*}{MNIST}
& \multirow{4}{*}{52}  
& 1   & 12.56 & 0.8462 & 12.53 & 0.8469 \\
&                       & 2    & 8.623 & 0.9255 & 8.588 & 0.9261 \\
&                       & 4    & 6.957 & 0.9512 & 6.934 & 0.9515 \\
&                       & FSL  & --- & --- & 12.46 & 0.8486 \\
\midrule
\multirow{4}{*}{CIFAR-B\&W}
& \multirow{4}{*}{154} 
& 1   & 3.476 & 0.9877 & 3.479 & 0.9877 \\
&                       & 2    & 3.232 & 0.9893 & 3.228 & 0.9894 \\
&                       & 4    & 2.867 & 0.9916 & 2.863 & 0.9916 \\
&                       & FSL  & --- & --- & 3.470 & 0.9877 \\
\midrule
\multirow{4}{*}{CIFAR-RGB}
& \multirow{4}{*}{154} 
& 1   & 3.575 & 0.9870 & 3.576 & 0.9870 \\
&                       & 2    & 3.303 & 0.9889 & 3.299 & 0.9889 \\
&                       & 4    & 3.034 & 0.9906 & 3.031 & 0.9906 \\
&                       & FSL  & --- & --- & 3.590 & 0.9870 \\
\bottomrule
\end{tabular}
\end{table*}

\section{Discussion}
We applied the adaptive interpolating quantum transform (AIQT) as a replacement for the Fourier transform in sparse amplitude encoding of real-world data. By learning a data-adaptive basis, the AIQT reduces the information lost in the coefficient truncation step, maximizing the information captured by the retained coefficients. On a one-dimensional finance dataset, the AIQT cut validation reconstruction error by about 40\% relative to the Fourier-based method at matched sparsity, while also improving fidelity. More importantly, for this dataset the AIQT scales more favorably with the sparsity budget: as $k$ increases, the approximation cRMSE decays more rapidly than for the FSL. Although the AIQT does not inherit the exact real-valued guarantee of the Fourier transform, our experiments show that the trained models naturally suppress imaginary leakage: the imaginary-part norm drops by several orders of magnitude during training, and the reconstructed states remain almost perfectly real-valued across sparsity levels. We also introduced a deep variant formed by stacking multiple AIQT blocks. This ''deep AIQT'' architecture further improves reconstruction quality on image data, achieving up to roughly 50\% reduction in error and higher fidelity compared to the Fourier-based method. These results suggest that the AIQT provides a practical, quantum-native way to improve approximate amplitude encoding by reducing reconstruction error and improving fidelity at fixed sparsity. Moreover, for the shallow AIQT, these improvements are achieved at a similar gate cost to the FSL. Future work could incorporate gate-aware regularization into the loss function to further reduce encoding cost.

Thus, our work provides a data-driven approach to addressing a central practical bottleneck in quantum computing: quantum state preparation from classical data. In particular, the parameters of the AIQT can be trained from unlabeled inputs entirely on a classical computer, avoiding quantum-hardware sampling during training. Moreover, the FFT-like butterfly structure of the AIQT yields a quasilinear classical forward cost in $N$ (i.e., $O(N\log N)$), making training far cheaper than learning a generic dense $N\times N$ transform. The parameter count is also dramatically reduced: whereas an unconstrained $N\times N$ unitary scales with $O(N^{2})$ parameters, the AIQT uses only $O(n^{2})$ trainable parameters for $N=2^{n}$. Once trained, the learned transform maps directly to a QFT-like quantum circuit with $O(n^{2})$ gates on an $n$-qubit system. More broadly, the deep AIQT architecture increases expressivity and motivates exploring applications beyond sparse amplitude encoding.

\section{Methods}
\subsection{The AIQT implementation, initialization, and training}
We now describe the AIQT implementation used throughout this work. The AIQT is a trainable unitary transform with an FFT/QFT-like butterfly structure \cite{nielsen2010,butterfly_1,butterfly_2}. For $N=2^{n}$, the computation proceeds in $n$ stages. At each stage, the vector is split into even and odd subsequences; the odd branch receives a trainable phase modulation that corresponds to controlled-phase operations in the AIQT circuit. The two branches are then mixed by a trainable $2\times 2$ unitary parameterized as $\mathrm{U3}(\alpha,\beta,\gamma)$ (Fig.~\ref{fig1}c). The trainable parameters are the per-stage mixer angles $(\boldsymbol{\alpha},\boldsymbol{\beta},\boldsymbol{\gamma})$ and the controlled-phase angles $\boldsymbol{\theta} \in \mathbb{R}^{n(n-1)/2}$.

We initialize the AIQT parameters to reproduce the quantum Fourier transform. Specifically, we set all mixer gates to $\mathrm{U3}(\tfrac{\pi}{2},0,\pi)$ and initialize the controlled-phase gates as $\mathrm{CR}(\theta)=\mathrm{diag}(1,1,1,e^{i\theta})$ with phase angles
\begin{equation}
\boldsymbol{\theta}^{(L)}_{\mathrm{init}} =
\bigl(-\tfrac{\pi}{2^{L}}, -\tfrac{\pi}{2^{L-1}}, \dots, -\tfrac{\pi}{2}\bigr),
\quad L = 1,\dots,n-1,
\end{equation}
which we pack as
\begin{equation}
\boldsymbol{\theta}_{\mathrm{Fourier}}
=
\bigl(\boldsymbol{\theta}^{(1)}_{\mathrm{init}},\dots,\boldsymbol{\theta}^{(n-1)}_{\mathrm{init}}\bigr).
\end{equation}
Here, $L$ indexes the phase-ladder levels, of which there are $n-1$ for an $n$-qubit system.

To minimize $\mathcal{L}_{\text{tail}}$ using automatic differentiation, we address the non-differentiable nature of the discrete coefficient selection $\mathcal{K}(x)$. We therefore optimize a differentiable relaxation of the loss using a sigmoid-based soft mask (temperature $\tau = 10^{-2}$) computed from coefficient energies relative to the $k$-th largest threshold, with gradients propagated via a straight-through estimator. To improve convergence stability and discourage degenerate uniform-energy solutions, we add a small entropy regularization term ($\lambda = 10^{-4}$) to the objective function.

We optimize the parameters of the AIQT using Adam for 150 epochs (batch size 128) and a cosine-annealing learning-rate schedule that decays over the first two-thirds of training and is held at its minimum value for the final third.

\subsection{Coefficient selection}
As described in the workflow in Section~\nameref{sec:workflow}, given coefficients $\mathbf{y}(\mathbf{x})$, we form an index set $\mathcal{K}(\mathbf{x})$ that specifies which coefficients are retained in the top-$k$ truncation step. In this subsection, we describe how $\mathcal{K}(\mathbf{x})$ is constructed for the AIQT and FSL workflows.

For coefficient selection in the AIQT workflow, we construct $\mathcal{K}(\mathbf{x})$ by selecting the indices of the $k$ largest coefficients of $\mathbf{y}(\mathbf{x})$ in squared magnitude. For the FSL baseline, coefficient selection must preserve the conjugate symmetry that arises in Fourier coefficients of real-valued inputs. Accordingly, our truncation step explicitly enforces conjugate symmetry when constructing $\mathcal{K}(\mathbf{x})$ as follows. We first include the first coefficient $j = 1$ and, for even $N$, the middle coefficient $j = N/2 + 1$ in $\mathcal{K}(\mathbf{x})$. These are always retained and do not count against the top-$k$ budget. We then form conjugate-symmetric pairs $(j, N+2-j)$ for $j = 2,\dots, N/2 $ and rank these pairs by $m_j(\mathbf{x}) + m_{N+2-j}(\mathbf{x})$, with $m_j(\mathbf{x})$ defined in Eq.~\ref{eq:mj}. From this ordering, we select the top $k/2$ pairs and include both indices $j$ and $N+2-j$  from each selected pair in $\mathcal{K}(\mathbf{x})$. Thus, a top-$k$ budget corresponds to $k$ coefficients drawn from conjugate-symmetric pairs and, together with the first and middle coefficients, yields $\lvert\mathcal{K}(\mathbf{x})\rvert = k+2$.

\subsection{Datasets}
Motivated by future applications in quantum finance and quantum image processing, we evaluate the performance of our method and the FSL on three datasets: a 1-D financial time-series dataset and two 2-D image datasets (MNIST and CIFAR). The financial dataset comprises stock prices sampled every 2 minutes for 217 companies, sourced from Yahoo Finance \cite{yahoo_finance}. Each series is segmented into overlapping windows of length $N=1024$ with a stride of 128 points, yielding 8669 samples in total. We split these chronologically into 80\% training and 20\% validation. For MNIST and CIFAR, we use 50k images from each dataset, considering both CIFAR-RGB (color) and CIFAR-B\&W (grayscale), and apply an 80/20 training/validation split. Each image is resized to $32 \times 32$, resulting in an input dimension $N = 1024$.

\subsection{Evaluation metrics}
To compare the performance of the AIQT and the FSL, we use two metrics: the complex root-mean-squared error (cRMSE) and the fidelity ($F$). For each input $\mathbf{x}$, let $\ket{\psi(\mathbf{x})}$ denote the exact amplitude-encoded quantum state and $\ket{\tilde{\psi}(\mathbf{x})}$ the reconstructed quantum state. Writing their amplitude vectors as $\boldsymbol{\psi}(\mathbf{x}), \tilde{\boldsymbol{\psi}}(\mathbf{x}) \in \mathbb{C}^N$ with components $\psi_j(\mathbf{x})$ and $\tilde{\psi}_j(\mathbf{x})$, we define the error components $e_j(\mathbf{x}) = \tilde{\psi}_j(\mathbf{x}) - \psi_j(\mathbf{x})$. The complex RMSE between $\ket{\psi(\mathbf{x})}$ and $\ket{\tilde{\psi}(\mathbf{x})}$ is then
\begin{equation}
    \operatorname{cRMSE}\bigl(\ket{\psi(\mathbf{x})},\ket{\tilde{\psi}(\mathbf{x})}\bigr)
    =
    \left(
        \frac{1}{N}
        \sum_{j=1}^N
        \bigl|e_j(\mathbf{x})\bigr|^2
    \right)^{1/2}.
\end{equation}
The fidelity between $\ket{\psi(\mathbf{x})}$ and $\ket{\tilde{\psi}(\mathbf{x})}$ is defined as
\begin{equation}
    F\bigl(\ket{\psi(\mathbf{x})},\ket{\tilde{\psi}(\mathbf{x})}\bigr)
    =
    \bigl|\langle \psi(\mathbf{x}) \mid \tilde{\psi}(\mathbf{x}) \rangle\bigr|^2,
\end{equation}
where $\langle \psi(\mathbf{x}) \mid \tilde{\psi}(\mathbf{x}) \rangle$ denotes the standard complex inner product. We report cRMSE and $F$ averaged over all samples in the dataset.

The entire workflow was implemented using the PyTorch framework.

\bmhead{Acknowledgements}
This work was supported by JSPS KAKENHI under Grant-in-Aid for Early-Career Scientists No. JP24K16985 and JSPS KAKENHI under Grant-in-Aid for Transformative Research Areas No. JP22H05114. This study was partially carried out using the facilities of the Supercomputer Center, the Institute for Solid State Physics, the University of Tokyo (ISSPkyodo-SC-2025-Eb-0008, 2025-Ea-0013) and the TSUBAME4.0 supercomputer at the Institute of Science Tokyo. This work was partially supported by the Center of Innovations for Sustainable Quantum AI (JST Grant Number JPMJPF2221). This work was supported by Japan Science and Technology Agency (JST) as part of Adopting Sustainable Partnerships for Innovative Research Ecosystem (ASPIRE), Grant Number JPMJAP24C1. The author acknowledges the contributions and discussions provided by the members of Quemix Inc.

\bibliography{manuscript}

@InProceedings{li_sparse_encode,
  author =	{Li, Lvzhou and Luo, Jingquan},
  title =	{{Nearly Optimal Circuit Size for Sparse Quantum State Preparation}},
  booktitle =	{52nd International Colloquium on Automata, Languages, and Programming (ICALP 2025)},
  pages =	{113:1--113:19},
  series =	{Leibniz International Proceedings in Informatics (LIPIcs)},
  ISBN =	{978-3-95977-372-0},
  ISSN =	{1868-8969},
  year =	{2025},
  volume =	{334},
  editor =	{Censor-Hillel, Keren and Grandoni, Fabrizio and Ouaknine, Jo\"{e}l and Puppis, Gabriele},
  publisher =	{Schloss Dagstuhl -- Leibniz-Zentrum f{\"u}r Informatik},
  address =	{Dagstuhl, Germany},
  URL =		{https://drops.dagstuhl.de/entities/document/10.4230/LIPIcs.ICALP.2025.113},
  URN =		{urn:nbn:de:0030-drops-234900},
  doi =		{10.4230/LIPIcs.ICALP.2025.113}
}

@article{mao_sparse_encode,
  title = {Toward optimal circuit size for sparse quantum state preparation},
  author = {Mao, Rui and Tian, Guojing and Sun, Xiaoming},
  journal = {Phys. Rev. A},
  volume = {110},
  issue = {3},
  pages = {032439},
  numpages = {9},
  year = {2024},
  month = {Sep},
  publisher = {American Physical Society},
  doi = {10.1103/PhysRevA.110.032439},
  url = {https://link.aps.org/doi/10.1103/PhysRevA.110.032439}
}

@article{sparse1,
  doi = {10.22331/q-2021-03-15-412},
  url = {https://doi.org/10.22331/q-2021-03-15-412},
  title = {Quantum {C}ircuits for {S}parse {I}sometries},
  author = {Malvetti, Emanuel and Iten, Raban and Colbeck, Roger},
  journal = {{Quantum}},
  issn = {2521-327X},
  publisher = {{Verein zur F{\"{o}}rderung des Open Access Publizierens in den Quantenwissenschaften}},
  volume = {5},
  pages = {412},
  month = mar,
  year = {2021}
}

@article{sparse2,
  title = {Double sparse quantum state preparation},
  volume = {21},
  ISSN = {1573-1332},
  url = {http://dx.doi.org/10.1007/s11128-022-03549-y},
  DOI = {10.1007/s11128-022-03549-y},
  number = {6},
  journal = {Quantum Information Processing},
  publisher = {Springer Science and Business Media LLC},
  author = {de Veras,  Tiago M. L. and da Silva,  Leon D. and da Silva,  Adenilton J.},
  year = {2022},
  month = jun 
}

@article{sparse3,
  title = {Efficient deterministic preparation of quantum states using decision diagrams},
  author = {Mozafari, Fereshte and De Micheli, Giovanni and Yang, Yuxiang},
  journal = {Phys. Rev. A},
  volume = {106},
  issue = {2},
  pages = {022617},
  numpages = {9},
  year = {2022},
  month = {Aug},
  publisher = {American Physical Society},
  doi = {10.1103/PhysRevA.106.022617},
  url = {https://link.aps.org/doi/10.1103/PhysRevA.106.022617}
}

@article{sparse4,
  title = {Simple quantum algorithm to efficiently prepare sparse states},
  author = {Ramacciotti, Debora and Lefterovici, Andreea I. and Rotundo, Antonio F.},
  journal = {Phys. Rev. A},
  volume = {110},
  issue = {3},
  pages = {032609},
  numpages = {10},
  year = {2024},
  month = {Sep},
  publisher = {American Physical Society},
  doi = {10.1103/PhysRevA.110.032609},
  url = {https://link.aps.org/doi/10.1103/PhysRevA.110.032609}
}

@article{sparse5,
  title = {Quantum State Preparation with Optimal Circuit Depth: Implementations and Applications},
  author = {Zhang, Xiao-Ming and Li, Tongyang and Yuan, Xiao},
  journal = {Phys. Rev. Lett.},
  volume = {129},
  issue = {23},
  pages = {230504},
  numpages = {6},
  year = {2022},
  month = {Nov},
  publisher = {American Physical Society},
  doi = {10.1103/PhysRevLett.129.230504},
  url = {https://link.aps.org/doi/10.1103/PhysRevLett.129.230504}
}

@article{topkFourier_theory,
doi = {10.1088/2058-9565/acfc62},
url = {https://doi.org/10.1088/2058-9565/acfc62},
year = {2023},
month = {oct},
publisher = {IOP Publishing},
volume = {9},
number = {1},
pages = {015002},
author = {Moosa, Mudassir and Watts, Thomas W and Chen, Yiyou and Sarma, Abhijat and McMahon, Peter L},
title = {Linear-depth quantum circuits for loading Fourier approximations of arbitrary functions},
journal = {Quantum Science and Technology}
}

@article{aiqt,
  title = {Adaptive interpolating quantum transform: A quantum-native framework for efficient transform learning},
  author = {Budiutama, Gekko and Daimon, Shunsuke and Nishi, Hirofumi and Kaneko, Ryui and Ohtsuki, Tomi and Matsushita, Yu-ichiro},
  journal = {Phys. Rev. A},
  volume = {112},
  issue = {6},
  pages = {062410},
  numpages = {7},
  year = {2025},
  month = {Dec},
  publisher = {American Physical Society},
  doi = {10.1103/vyr3-h9hq},
  url = {https://link.aps.org/doi/10.1103/vyr3-h9hq}
}

@misc{topkFourier_app,
      title={Quantum medical image encoding and compression using Fourier-based methods}, 
      author={Taehee Ko and Inho Lee and Hyeong Won Yu},
      year={2025},
      eprint={2505.06471},
      archivePrefix={arXiv},
      primaryClass={quant-ph},
      url={https://arxiv.org/abs/2505.06471}, 
}

@book{nielsen2010,
  title={Quantum Computation and Quantum Information: 10th Anniversary Edition},
  author={Nielsen, Michael A. and Chuang, Isaac L.},
  year={2010},
  publisher={Cambridge University Press},
  address={Cambridge, UK},
  isbn={978-1107002173}
}

@article{opening_2,
   author = "Kjaergaard, Morten and Schwartz, Mollie E. and Braumüller, Jochen and Krantz, Philip and Wang, Joel I.-J. and Gustavsson, Simon and Oliver, William D.",
   title = "Superconducting Qubits: Current State of Play", 
   journal= "Annual Review of Condensed Matter Physics",
   year = "2020",
   volume = "11",
   number = "Volume 11, 2020",
   pages = "369-395",
   doi = "https://doi.org/10.1146/annurev-conmatphys-031119-050605",
   url = "https://www.annualreviews.org/content/journals/10.1146/annurev-conmatphys-031119-050605",
   publisher = "Annual Reviews",
   issn = "1947-5462",
   type = "Journal Article",
  }

@article{opening_3,
  title = {Quantum computational chemistry},
  author = {McArdle, Sam and Endo, Suguru and Aspuru-Guzik, Al\'an and Benjamin, Simon C. and Yuan, Xiao},
  journal = {Rev. Mod. Phys.},
  volume = {92},
  issue = {1},
  pages = {015003},
  numpages = {51},
  year = {2020},
  month = {Mar},
  publisher = {American Physical Society},
  doi = {10.1103/RevModPhys.92.015003},
  url = {https://link.aps.org/doi/10.1103/RevModPhys.92.015003}
}

@article{opening_4,
author = {Endo ,Suguru and Cai ,Zhenyu and Benjamin ,Simon C. and Yuan ,Xiao},
title = {Hybrid Quantum-Classical Algorithms and Quantum Error Mitigation},
journal = {Journal of the Physical Society of Japan},
volume = {90},
number = {3},
pages = {032001},
year = {2021},
doi = {10.7566/JPSJ.90.032001},

URL = { 
    
        https://doi.org/10.7566/JPSJ.90.032001
    
    

},
eprint = { 
    
        https://doi.org/10.7566/JPSJ.90.032001
    
    

}
}

@article{opening_5,
  title = {Variational quantum algorithms},
  volume = {3},
  ISSN = {2522-5820},
  url = {http://dx.doi.org/10.1038/s42254-021-00348-9},
  DOI = {10.1038/s42254-021-00348-9},
  number = {9},
  journal = {Nature Reviews Physics},
  publisher = {Springer Science and Business Media LLC},
  author = {Cerezo,  M. and Arrasmith,  Andrew and Babbush,  Ryan and Benjamin,  Simon C. and Endo,  Suguru and Fujii,  Keisuke and McClean,  Jarrod R. and Mitarai,  Kosuke and Yuan,  Xiao and Cincio,  Lukasz and Coles,  Patrick J.},
  year = {2021},
  month = aug,
  pages = {625–644}
}

@article{opening_6,
  title = {Noisy intermediate-scale quantum algorithms},
  author = {Bharti, Kishor and Cervera-Lierta, Alba and Kyaw, Thi Ha and Haug, Tobias and Alperin-Lea, Sumner and Anand, Abhinav and Degroote, Matthias and Heimonen, Hermanni and Kottmann, Jakob S. and Menke, Tim and Mok, Wai-Keong and Sim, Sukin and Kwek, Leong-Chuan and Aspuru-Guzik, Al\'an},
  journal = {Rev. Mod. Phys.},
  volume = {94},
  issue = {1},
  pages = {015004},
  numpages = {69},
  year = {2022},
  month = {Feb},
  publisher = {American Physical Society},
  doi = {10.1103/RevModPhys.94.015004},
  url = {https://link.aps.org/doi/10.1103/RevModPhys.94.015004}
}

@article{opening_7,
  title = {Parameter estimation in quantum sensing based on deep reinforcement learning},
  volume = {8},
  ISSN = {2056-6387},
  url = {http://dx.doi.org/10.1038/s41534-021-00513-z},
  DOI = {10.1038/s41534-021-00513-z},
  number = {1},
  journal = {npj Quantum Information},
  publisher = {Springer Science and Business Media LLC},
  author = {Xiao,  Tailong and Fan,  Jianping and Zeng,  Guihua},
  year = {2022},
  month = jan 
}

@article{opening_8,
  title = {Practical advantage of quantum machine learning in ghost imaging},
  volume = {6},
  ISSN = {2399-3650},
  url = {http://dx.doi.org/10.1038/s42005-023-01290-1},
  DOI = {10.1038/s42005-023-01290-1},
  number = {1},
  journal = {Communications Physics},
  publisher = {Springer Science and Business Media LLC},
  author = {Xiao,  Tailong and Zhai,  Xinliang and Wu,  Xiaoyan and Fan,  Jianping and Zeng,  Guihua},
  year = {2023},
  month = jul 
}

@article{ae_1,
  title = {Synthesis of Quantum Superpositions by Quantum Computation},
  author = {Grover, Lov K.},
  journal = {Phys. Rev. Lett.},
  volume = {85},
  issue = {6},
  pages = {1334--1337},
  numpages = {0},
  year = {2000},
  month = {Aug},
  publisher = {American Physical Society},
  doi = {10.1103/PhysRevLett.85.1334},
  url = {https://link.aps.org/doi/10.1103/PhysRevLett.85.1334}
}

@article{ae_2,
  title = {Efficient scheme for initializing a quantum register with an arbitrary superposed state},
  author = {Long, Gui-Lu and Sun, Yang},
  journal = {Phys. Rev. A},
  volume = {64},
  issue = {1},
  pages = {014303},
  numpages = {4},
  year = {2001},
  month = {Jun},
  publisher = {American Physical Society},
  doi = {10.1103/PhysRevA.64.014303},
  url = {https://link.aps.org/doi/10.1103/PhysRevA.64.014303}
}

@article{ae_3,
  title = {Quantum circuits for incompletely specified two-qubit operators},
  volume = {5},
  ISSN = {1533-7146},
  url = {http://dx.doi.org/10.26421/QIC5.1-5},
  DOI = {10.26421/qic5.1-5},
  number = {1},
  journal = {Quantum Information and Computation},
  publisher = {Rinton Press},
  author = {Shende,  V.V. and Markov,  I.L.},
  year = {2005},
  month = jan,
  pages = {48–56}
}

@article{ae_4,
author = {M\"{o}tt\"{o}nen, Mikko and Vartiainen, Juha J. and Bergholm, Ville and Salomaa, Martti M.},
title = {Transformation of quantum states using uniformly controlled rotations},
year = {2005},
issue_date = {September 2005},
publisher = {Rinton Press, Incorporated},
address = {Paramus, NJ},
volume = {5},
number = {6},
issn = {1533-7146},
journal = {Quantum Info. Comput.},
month = sep,
pages = {467–473},
numpages = {7}
}

@article{ae_5,
  title = {Quantum-state preparation with universal gate decompositions},
  author = {Plesch, Martin and Brukner, \ifmmode \check{C}\else \v{C}\fi{}aslav},
  journal = {Phys. Rev. A},
  volume = {83},
  issue = {3},
  pages = {032302},
  numpages = {5},
  year = {2011},
  month = {Mar},
  publisher = {American Physical Society},
  doi = {10.1103/PhysRevA.83.032302},
  url = {https://link.aps.org/doi/10.1103/PhysRevA.83.032302}
}

@article{ae_6,
  title = {Low-depth quantum state preparation},
  author = {Zhang, Xiao-Ming and Yung, Man-Hong and Yuan, Xiao},
  journal = {Phys. Rev. Res.},
  volume = {3},
  issue = {4},
  pages = {043200},
  numpages = {14},
  year = {2021},
  month = {Dec},
  publisher = {American Physical Society},
  doi = {10.1103/PhysRevResearch.3.043200},
  url = {https://link.aps.org/doi/10.1103/PhysRevResearch.3.043200}
}

@article{ae_7,
  title = {Efficient state preparation for a register of quantum bits},
  author = {Soklakov, Andrei N. and Schack, R\"udiger},
  journal = {Phys. Rev. A},
  volume = {73},
  issue = {1},
  pages = {012307},
  numpages = {13},
  year = {2006},
  month = {Jan},
  publisher = {American Physical Society},
  doi = {10.1103/PhysRevA.73.012307},
  url = {https://link.aps.org/doi/10.1103/PhysRevA.73.012307}
}

@article{ae_8,
  title = {Efficient quantum algorithm for preparing molecular-system-like states on a quantum computer},
  author = {Wang, Hefeng and Ashhab, S. and Nori, Franco},
  journal = {Phys. Rev. A},
  volume = {79},
  issue = {4},
  pages = {042335},
  numpages = {9},
  year = {2009},
  month = {Apr},
  publisher = {American Physical Society},
  doi = {10.1103/PhysRevA.79.042335},
  url = {https://link.aps.org/doi/10.1103/PhysRevA.79.042335}
}

@article{ae_9,
  title = {Circuit-Based Quantum Random Access Memory for Classical Data},
  volume = {9},
  ISSN = {2045-2322},
  url = {http://dx.doi.org/10.1038/s41598-019-40439-3},
  DOI = {10.1038/s41598-019-40439-3},
  number = {1},
  journal = {Scientific Reports},
  publisher = {Springer Science and Business Media LLC},
  author = {Park,  Daniel K. and Petruccione,  Francesco and Rhee,  June-Koo Kevin},
  year = {2019},
  month = mar 
}

@article{ae_10,
  title = {Quantum state preparation protocol for encoding classical data into the amplitudes of a quantum information processing register's wave function},
  author = {Ashhab, Sahel},
  journal = {Phys. Rev. Res.},
  volume = {4},
  issue = {1},
  pages = {013091},
  numpages = {7},
  year = {2022},
  month = {Feb},
  publisher = {American Physical Society},
  doi = {10.1103/PhysRevResearch.4.013091},
  url = {https://link.aps.org/doi/10.1103/PhysRevResearch.4.013091}
}

@article{ae_11,
  title = {Qubit encoding for a mixture of localized functions},
  author = {Kosugi, Taichi and Daimon, Shunsuke and Nishi, Hirofumi and Tsuneyuki, Shinji and Matsushita, Yu-ichiro},
  journal = {Phys. Rev. A},
  volume = {110},
  issue = {6},
  pages = {062407},
  numpages = {18},
  year = {2024},
  month = {Dec},
  publisher = {American Physical Society},
  doi = {10.1103/PhysRevA.110.062407},
  url = {https://link.aps.org/doi/10.1103/PhysRevA.110.062407}
}

@article{budiutama2024,
  title = {Channel attention for quantum convolutional neural networks},
  author = {Budiutama, Gekko and Daimon, Shunsuke and Nishi, Hirofumi and Kaneko, Ryui and Ohtsuki, Tomi and Matsushita, Yu-ichiro},
  journal = {Phys. Rev. A},
  volume = {110},
  issue = {1},
  pages = {012447},
  numpages = {9},
  year = {2024},
  month = {Jul},
  publisher = {American Physical Society},
  doi = {10.1103/PhysRevA.110.012447},
  url = {https://link.aps.org/doi/10.1103/PhysRevA.110.012447}
}

@misc{gt,
      title={General Transform: A Unified Framework for Adaptive Transform to Enhance Representations}, 
      author={Gekko Budiutama and Shunsuke Daimon and Hirofumi Nishi and Yu-ichiro Matsushita},
      year={2025},
      eprint={2505.04969},
      archivePrefix={arXiv},
      primaryClass={cs.LG},
      url={https://arxiv.org/abs/2505.04969}, 
}

@article{ae_12,
  title = {Quantum circuit generation for amplitude encoding using a transformer decoder},
  author = {Daimon, Shunsuke and Matsushita, Yu-ichiro},
  journal = {Phys. Rev. Appl.},
  volume = {22},
  issue = {4},
  pages = {L041001},
  numpages = {6},
  year = {2024},
  month = {Oct},
  publisher = {American Physical Society},
  doi = {10.1103/PhysRevApplied.22.L041001},
  url = {https://link.aps.org/doi/10.1103/PhysRevApplied.22.L041001}
}

@misc{yahoo_finance,
  title        = {Yahoo Finance: Historical Market Data},
  howpublished = {\url{https://finance.yahoo.com/}},
}

@misc{hwsl,
      title={Hybrid Quantum State Preparation via Data Compression}, 
      author={Emad Rezaei Fard Boosari and Maryam Afsary},
      year={2025},
      eprint={2512.01798},
      archivePrefix={arXiv},
      primaryClass={quant-ph},
      url={https://arxiv.org/abs/2512.01798}, 
}

@article{ae_13,
  title = {Quantum Generative Adversarial Networks for learning and loading random distributions},
  volume = {5},
  ISSN = {2056-6387},
  url = {http://dx.doi.org/10.1038/s41534-019-0223-2},
  DOI = {10.1038/s41534-019-0223-2},
  number = {1},
  journal = {npj Quantum Information},
  publisher = {Springer Science and Business Media LLC},
  author = {Zoufal,  Christa and Lucchi,  Aurélien and Woerner,  Stefan},
  year = {2019},
  month = nov 
}

@article{ae_14,
  title = {Approximate amplitude encoding in shallow parameterized quantum circuits and its application to financial market indicators},
  author = {Nakaji, Kouhei and Uno, Shumpei and Suzuki, Yohichi and Raymond, Rudy and Onodera, Tamiya and Tanaka, Tomoki and Tezuka, Hiroyuki and Mitsuda, Naoki and Yamamoto, Naoki},
  journal = {Phys. Rev. Res.},
  volume = {4},
  issue = {2},
  pages = {023136},
  numpages = {20},
  year = {2022},
  month = {May},
  publisher = {American Physical Society},
  doi = {10.1103/PhysRevResearch.4.023136},
  url = {https://link.aps.org/doi/10.1103/PhysRevResearch.4.023136}
}

@article{ae_15,
  title = {Approximate complex amplitude encoding algorithm and its application to data classification problems},
  author = {Mitsuda, Naoki and Ichimura, Tatsuhiro and Nakaji, Kouhei and Suzuki, Yohichi and Tanaka, Tomoki and Raymond, Rudy and Tezuka, Hiroyuki and Onodera, Tamiya and Yamamoto, Naoki},
  journal = {Phys. Rev. A},
  volume = {109},
  issue = {5},
  pages = {052423},
  numpages = {12},
  year = {2024},
  month = {May},
  publisher = {American Physical Society},
  doi = {10.1103/PhysRevA.109.052423},
  url = {https://link.aps.org/doi/10.1103/PhysRevA.109.052423}
}

@article{butterfly_1,
 ISSN = {00255718, 10886842},
 URL = {http://www.jstor.org/stable/2003354},
 author = {James W. Cooley and John W. Tukey},
 journal = {Mathematics of Computation},
 number = {90},
 pages = {297--301},
 publisher = {American Mathematical Society},
 title = {An Algorithm for the Machine Calculation of Complex Fourier Series},
 urldate = {2026-01-04},
 volume = {19},
 year = {1965}
}

@inproceedings{butterfly_2,
  title     = {Learning Fast Algorithms for Linear Transforms Using Butterfly Factorizations},
  author    = {Dao, Tri and Gu, Albert and Eichhorn, Matthew and Rudra, Atri and R{\'e}, Christopher},
  booktitle = {Proceedings of the 36th International Conference on Machine Learning},
  series    = {PMLR},
  volume    = {97},
  pages     = {1517--1527},
  year      = {2019},
  editor    = {Chaudhuri, Kamalika and Salakhutdinov, Ruslan}
}

@misc{ae_16,
      title={Efficient quantum state preparation of multivariate functions using tensor networks}, 
      author={Marco Ballarin and Juan José García-Ripoll and David Hayes and Michael Lubasch},
      year={2025},
      eprint={2511.15674},
      archivePrefix={arXiv},
      primaryClass={quant-ph},
      url={https://arxiv.org/abs/2511.15674}, 
}
\end{document}